\documentclass[aps,prb,letterpaper,twocolumn]{revtex4}

\usepackage{graphicx}

\renewcommand{\deg}{$^\circ$}

\begin{document}

\title{Nucleation and Growth of GaN/AlN Quantum Dots}

\author{C. Adelmann}
\altaffiliation{Present address: Department of Chemical
Engineering and Materials Science, University of Minnesota, Minneapolis, Minnesota 55455-0132, USA.}
\author{B. Daudin}
\affiliation{CEA/CNRS Research Group ``Nanophysique et Semiconducteurs'', D\'epartement de Recherche Fondamentale sur la Mati\`ere Condens\'ee, CEA/Grenoble, 17 Rue des Martyrs, 38054 Grenoble Cedex 9, France}

\author{R. A. Oliver}
\altaffiliation{Present address: Department of Materials Science and Metallurgy, University of Cambridge, Pembroke Street, Cambridge CB2 3QZ, United Kingdom.}
\author{G. A. D. Briggs}
\affiliation{Department of Materials, University of Oxford, Parks Road, Oxford OX1 3PH, United Kingdom}

\author{R. E. Rudd}
\affiliation{Lawrence Livermore National Laboratory, 7000 East Avenue, Livermore, California 94550, USA}

\begin{abstract}

We study the nucleation of GaN islands grown by plasma-assisted molecular-beam epitaxy on AlN(0001) in a Stranski-Krastanov mode. In particular, we assess the variation of their height and density as a function of GaN coverage. We show that the GaN growth passes four stages: initially, the growth is layer-by-layer; subsequently, two-dimensional precursor islands form, which transform into genuine three-dimensional islands. During the latter stage, island height and density increase with GaN coverage until the density saturates. During further GaN growth, the density remains constant and a bimodal height distribution appears. The variation of island height and density as a function of substrate temperature is discussed in the framework of an equilibrium model for Stranski-Krastanov growth.

\end{abstract}

\pacs{81.07.Ta; 68.37.Ps; 81.10.Aj; 68.55.Ac}

\maketitle

\section{Introduction}

Zero-dimensional semiconductor quantum dots (QDs) have attracted much interest in the last decade due to their multiple potential applications ranging from low-threshold lasers\cite{Arakawa,Bimberg} via single-electron tunneling devices\cite{Odintsov,Kouwenhoven} to possible realizations of qubits for quantum computation.\cite{QC1,QC2} A versatile method for the fabrication of semiconductor QDs is their self-assembled growth following the Stranski-Krastanov (SK) growth mode.\cite{ReviewsSK} This mode usually occurs during the growth of semiconductor epilayers under compressive strain. Example material systems are In$_x$Ga$_{1-x}$As/GaAs,\cite{Goldstein,Leonard93,Moison} Si$_x$Ge$_{1-x}$/Si,\cite{Eaglesham,Mo} CdSe/ZnSe,\cite{Xin} or GaN/AlN.\cite{Daudin,Widmann} In this mode, atoms are initially deposited in form of a two-dimensional pseudomorphic wetting layer. The associated strain energy increases with the thickness of the wetting layer and is finally elastically relieved by the formation of islands. 

The usefulness of such self-assembled nanostructures relies on the ability to obtain homogeneous size distributions as well as to control their size, density and position. Many theoretical contributions have enhanced our understanding of the size distributions of SK-grown islands, but some controversy remains.\cite{Shchukin,Daruka,Williams,Priester,Chen,Ross,Jesson} Also, despite many experimental studies, the influence of growth parameters on the size and density of such islands is not fully understood, owing to the complexity of the physics of strained layer growth. 

In this work, we present results on the nucleation of GaN islands on AlN following an SK mode, in particular on the dependence on the amount of deposited GaN and the substrate temperature. We find that the qualitative behavior is similar to that found in other systems, e.g. for InAs/GaAs and Ge/Si. This reinforces the idea that there are common features of semiconductor SK growth, which are rather universal and independent of the specific material system. Absolute island sizes then densities will depend on the material system, possibly through material parameters like lattice misfit, elastic constants, or surface energies. Hence, we may further discuss the experimental data in the framework of an equilibrium model based on statistical physics that was originally developed in work on the Ge/Si system.\cite{Rudd}

\section{Experimental}

The samples have been grown in a MECA2000 molecular-beam epitaxy (MBE) chamber equipped with standard effusion cells for Ga and Al evaporation. The chamber also contains an rf plasma cell to provide active nitrogen for GaN and AlN growth. The pseudo-substrates used were about 2\,$\mu$m thick GaN(0001) (Ga-polarity) layers grown by metal-organic chemical vapor deposition on sapphire. The substrate temperature $T_S$ was measured by a thermocouple in mechanical contact to the backside of the molybdenum sample holder. To ensure substrate temperature reproducibility, each series of samples described below was grown on a single molybdenum substrate holder. 

\begin{figure}[tb]
\includegraphics[width=6cm,clip]{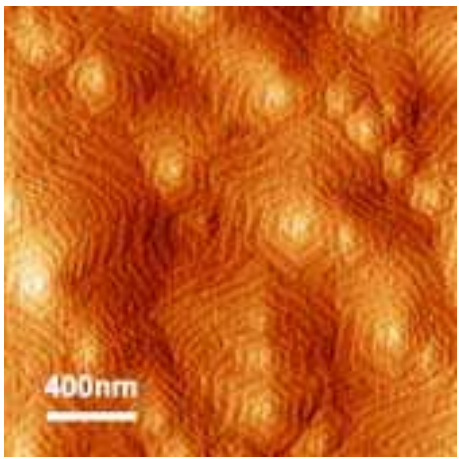}
\caption{\label{Fig:AFMAlN} \small Atomic-force micrograph of an AlN pseudosubstrate.}
\end{figure}

Prior to all experiments, a 100 nm thick GaN layer was grown under Ga-rich conditions on the pseudosubstrates to avoid the influence of a possible surface contamination layer. Subsequently, a 300\,nm thick AlN film was deposited under Al-rich conditions at a substrate temperature of 730\,\deg C. We have found by reflection high-energy electron diffraction (RHEED) and high-resolution X-Ray diffraction that this thickness is sufficient for the AlN layer to be virtually fully relaxed with a residual in-plane strain of $\epsilon_1 < 0.1$\%.\cite{Bellet} An atomic-force microscopy (AFM) image of such an AlN(0001) surface is shown in Fig.~\ref{Fig:AFMAlN}. The surface is characterized by about 30\,nm wide terraces and spiral hillocks, similar to GaN surfaces grown under equivalent conditions.\cite{AdelmannJAP}

The growth rate and the GaN coverage have been experimentally determined for each sample by RHEED oscillations occurring during the growth of the wetting layer prior to island formation.\cite{Mula} Typically, growth rates for different layers (and different samples) were reproducible within about 0.01\,monolayers (ML)/s. The GaN coverage was then calculated with a precision better than 0.1\,ML. It is worth noting that no measurable influence of the growth rate on GaN island properties has been found in the range between 0.1 and 0.6\,ML/s. As, on the other hand, the Ga/N flux ratio has been found to have a crucial influence on the growth mode,\cite{Mula} it was fixed to 0.8 (N-rich conditions), which leads to a critical thickness for the SK transition of 2.25\,ML.\cite{Mula,AdelmannThesis} The critical thickness was measured for each sample and found to be reproducible within 0.1\,ML.

To study the GaN islands \emph{as grown}, the samples were rapidly quenched to room temperature under an N-flux. \emph{Ex-situ} AFM was then used to study the GaN morphology after exposure of the samples to air.

\begin{figure}[b]
\includegraphics[width=8cm,clip]{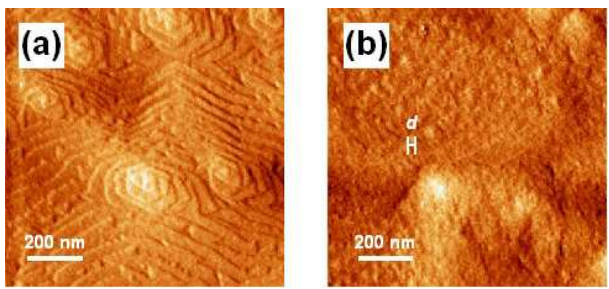}
\caption{\label{Fig:AFMQDfTheta} \small Atomic-force micrograph of the GaN surface morphology for coverages $\Theta$ of \textbf{(a)} 1.8\,ML and \textbf{(b)} 2.2\,ML, respectively. In (b), $d \approx 20$\,nm indicates the typical lateral length scale of the 2D precursor islands.}
\end{figure}

\section{Results}

\subsection{Dependence on GaN coverage}

The parameter most directly governing the properties of GaN islands grown in an SK mode on AlN is the GaN coverage $\Theta$, which effectively describes the time evolution of the islands during growth. To study this evolution, a series of samples has been grown at a substrate temperature of 730\,\deg C and a growth rate of 0.15\,ML/s. The GaN coverage was varied between $\Theta = 1.8$\,ML and $\Theta = 4.6$\,ML.

Figure \ref{Fig:AFMQDfTheta}(a) shows an AFM image of the morphology of GaN layers obtained after the deposition of 1.8\,ML, \emph{i.e.} for a coverage well below the critical thickness of 2.25\,ML. We find that the morphology is unchanged with respect to that of the AlN pseudosubstrate: the surface is characterized by about 30\,nm wide terraces and spiral hillocks. We can thus infer that the growth of about the first 2\,ML of GaN occurs in a layer-by-layer mode since RHEED oscillations are observed.\cite{Mula}

\begin{figure}[tb]
\includegraphics[width=8cm,clip]{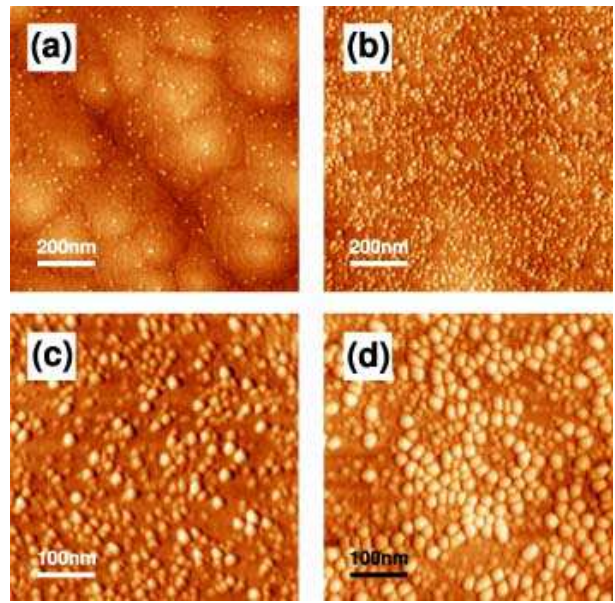}
\caption{\label{Fig:AFMQDfTheta2} \small Atomic-force micrograph of GaN surface morphology for coverages $\Theta$ of \textbf{(a)} 2.5\,ML, \textbf{(b)} 2.8\,ML, \textbf{(c)} 3.0\,ML, and \textbf{(d)} 4.6\,ML, respectively.}
\end{figure}

When the second monolayer is completed, the morphology changes, as evidenced in Fig.~\ref{Fig:AFMQDfTheta}(b) for $\Theta = 2.2$\,ML, \emph{i.e.} immediately before the SK transition. Remainders of terraces and spiral hillocks are still visible, but the surface is characterized on a short scale by 1--2\,ML high flat 2D islands with typical diameters of $d \approx 20$\,nm (see Fig.~\ref{Fig:AFMQDfTheta}(b)). The behavior can thus be described by a transition from a layer-by-layer growth to multilayer growth at around 2.0\,ML. Such an occurrence of 2D precursor islands prior to the genuine 2D-3D SK transition has also been observed in the InAs/(Al,Ga)As system.\cite{Cirlin,Kitabayashi,Ramachandran,Ballet}

When the growth is continued, the genuine 2D-3D transition occurs and 3D GaN islands are formed as shown in Fig.~\ref{Fig:AFMQDfTheta2}. We find that their density increases strongly with GaN coverage and saturates around 3.0\,ML at a value of $1.8\times 10^{11}$\,cm$^{-2}$ [see Figs.~\ref{Fig:AFMQDfTheta2}(a) and (b)]. Further GaN deposition does not lead to an increase in island density but instead islands grow in size. However, the islands do not grow continuously in size but a bimodal size distribution is observed [see Figs.~\ref{Fig:AFMQDfTheta2}(c) and (d)]. 

\begin{figure}[tb]
\includegraphics[width=8cm,clip]{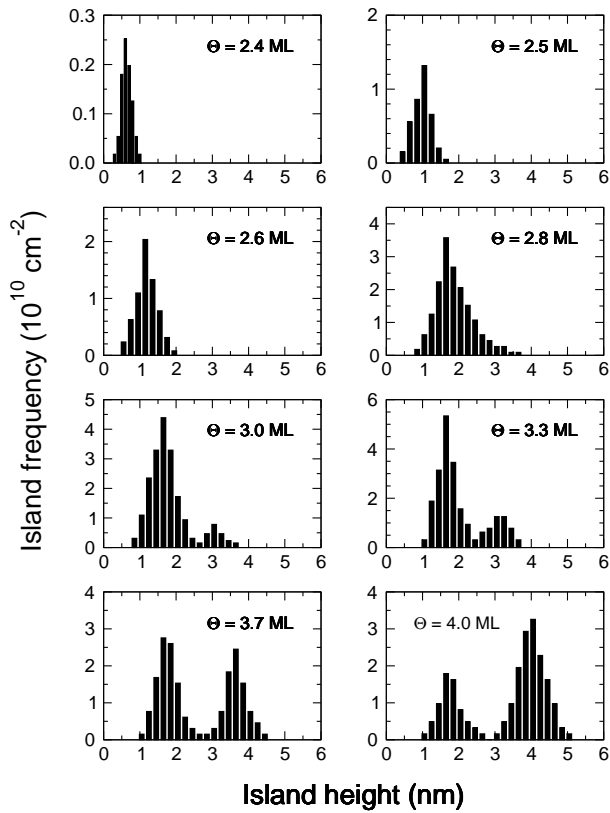}
\caption{\label{Fig:StatQDfTheta} \small Height distribution of GaN islands grown on AlN with nominal coverages as indicated. $T_S = 730$\,\deg C.}
\end{figure}

\begin{figure}[t]
\includegraphics[width=8cm,clip]{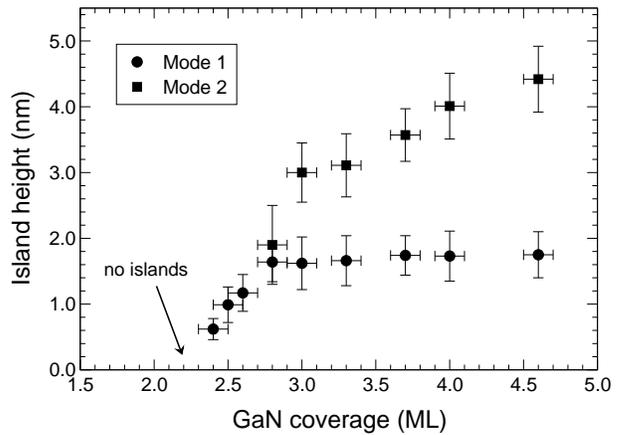}
\caption{\label{Fig:HeightQDfTheta} \small Average island height as a function of GaN coverage at $T_S = 730$\,\deg C. Above 2.8\,ML a bimodal size distribution is observed.}
\end{figure}

The variation of the island size is summarized in Figs.~\ref{Fig:StatQDfTheta} and \ref{Fig:HeightQDfTheta}. We see that, initially, the islands' height increases from 0.7\,nm to 1.6\,nm between 2.4 and 2.8\,ML of GaN coverage. For higher GaN coverage, a bimodal size distribution is observed with the average height of the first mode islands remaining constant at 1.6 nm independent of coverage. In contrast, the average height corresponding to the second mode increases continuously reaching 4.2\,nm at 4.7\,ML coverage with no sign of saturation in the examined coverage range. Differences in the shapes of mode 1 and mode 2 islands will be discussed below.

\begin{figure}[b]
\includegraphics[width=8cm,clip]{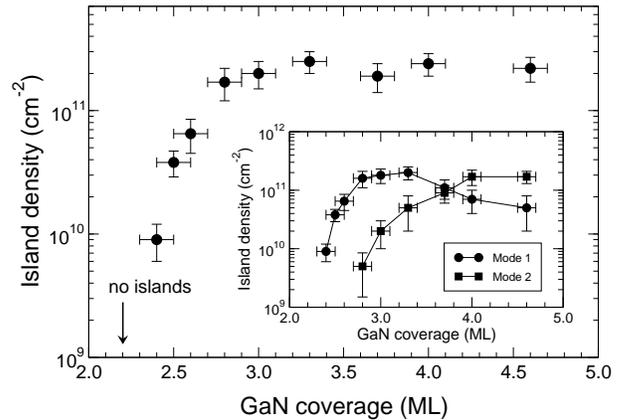}
\caption{\label{Fig:RhoQDfTheta} \small Total GaN island density as a function of GaN coverage at $T_S = 730$\,\deg C. The inset shows the partial density of the islands in the two modes. We observe that mode 1 islands transform into mode 2 islands after the deposition of about 2.8 -- 3.0\,ML.}
\end{figure}

The variation of the island density is depicted in Fig.~\ref{Fig:RhoQDfTheta}. We observe that the total density increases strongly after the 2D--3D transition but saturates after 2.8\,ML, i.e. at the coverage where the island size distribution becomes bimodal. The partial densities of the two modes are shown in the inset in Fig.~\ref{Fig:RhoQDfTheta}. For coverages below 2.8\,ML, the density of mode 1 islands is identical with the total density and increases strongly with coverage. After the deposition of 2.8\,ML, the density of mode 2 islands increases strongly, similarly to the behavior of mode 1 islands after the 2D--3D transition, whilst the density of mode 1 islands decreases. As the total island density remains approximately constant, mode 1 islands transform into mode 2 islands, probably without additional nucleation of new (mode 1) islands.

\subsection{Dependence on substrate temperature}

The influence of the substrate temperature is studied in a series of samples with $\Theta = 3.0$\,ML deposited at substrate temperatures between $T_S = 690$\,\deg C and $T_S =  760$\,\deg C. At lower temperatures, no SK growth is observed and GaN grows in a pseudo-2D mode,\cite{Daudin,Mula,AdelmannThesis} probably due to excessively low adatom mobility.\cite{Zywietz} Higher substrate temperatures are prohibited by our experimental apparatus due to indium bonding of the substrates to the substrate holder. All samples in this series were grown at a growth rate of 0.25\,ML/s.

\begin{figure}[tb]
\includegraphics[width=8cm,clip]{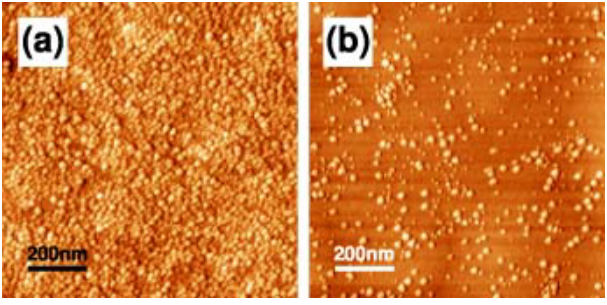}
\caption{\label{Fig:AFMQDfTS} \small Atomic-force micrograph of GaN surface morphology as a function of substrate temperature: \textbf{(a)} $T_S = 700$\,\deg C; \textbf{(a)} $T_S = 760$\,\deg C.}
\end{figure}

\begin{figure}[b]
\includegraphics[width=8cm,clip]{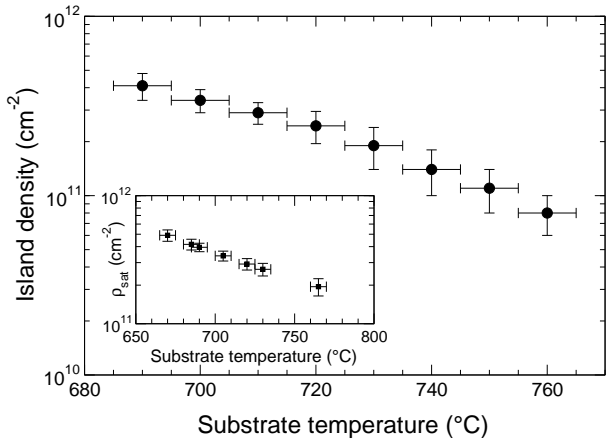}
\caption{\label{Fig:RhoQDfTS} \small GaN island density $\rho$ for a coverage of $\Theta = 3.0$\,ML as a function of substrate temperature $T_S$. The inset shows the island saturation density as a function of substrate temperature.}
\end{figure}

An AFM image of the morphology of two GaN layers grown at substrate temperatures of $T_S = 700$\,\deg C and $T_S = 760$\,\deg C, respectively, are shown in Fig.~\ref{Fig:AFMQDfTS}. We observe that the total island density decreases rapidly with substrate temperature, in keeping with previous results\cite{Widmann} and also with results obtained for the InAs/GaAs\cite{Leon1} and Ge/Si\cite{Kamins} systems. The variation of the GaN island density as a function of substrate temperature for a coverage of $\Theta = 3.0$\,ML is depicted in Fig.~\ref{Fig:RhoQDfTS}. We find an approximately exponential decrease of the island density with increasing substrate temperature from $4.1\times 10^{11}$\,cm$^{-2}$ at $T_S = 690$\,\deg C to  $8.0\times 10^{10}$\,cm$^{-2}$ at $T_S = 760$\,\deg C. 

\begin{figure}[tb]
\includegraphics[width=8cm,clip]{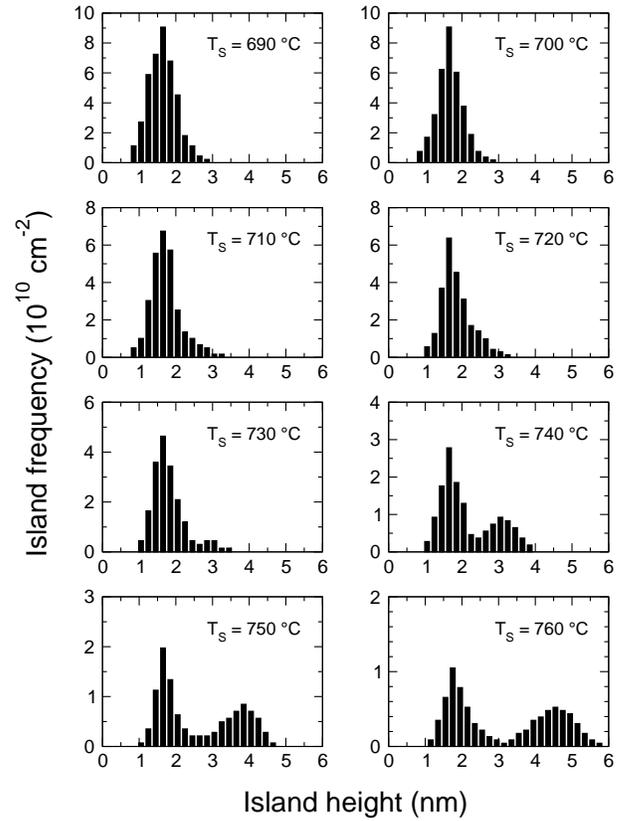}
\caption{\label{Fig:StatQDfTS} \small Height distribution of GaN islands grown on AlN at substrate temperatures as indicated. $\Theta = 3.0$\,ML.}
\end{figure}

\begin{figure}[tb]
\includegraphics[width=8cm,clip]{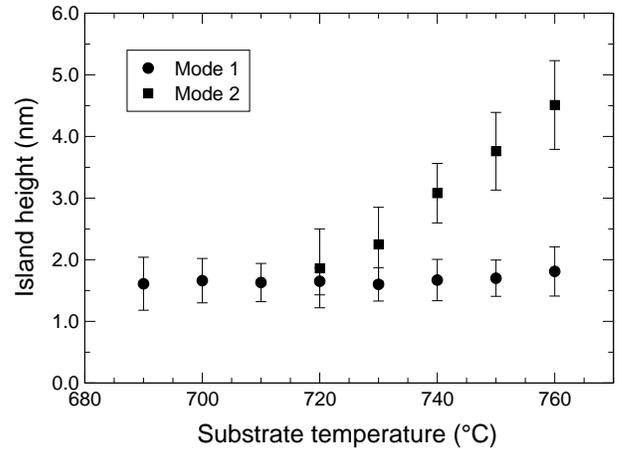}
\caption{\label{Fig:HeightQDfTS} \small Average island height as a function of substrate temperature for a GaN coverage of $\Theta = 3.0$\,ML. Above 720\,\deg C a bimodal size distribution is observed.}
\end{figure}

As we have seen above, the island density tends to saturate at sufficiently high GaN coverage. This saturation density is shown as a function of substrate temperature in the inset in Fig.~\ref{Fig:RhoQDfTS}. For substrate temperatures of $T_S \le 730$\,\deg C, the saturation density is similar to the density for $\Theta = 3.0$\,ML, in agreement with the results in the last section. However, for higher substrate temperature, the saturation density is significantly higher, indicating that saturation occurs for GaN coverages larger than 3.0\,ML.

Figure \ref{Fig:StatQDfTS} shows the island height distribution for $\Theta = 3.0$\,ML and substrate temperatures as indicated. The behavior is summarized in Fig.~\ref{Fig:HeightQDfTS}. We find a single approximately Gaussian distribution (in the limit of the statistical precision) at a height of 1.6\,nm for substrate temperatures $T_S \le 720$\,\deg C. These islands correspond to the mode 1 islands observed at $T_S = 730$\,\deg C and discussed above. At low temperature, the distribution is thus still monomodal after 3.0\,ML. For $T_S = 730$\,\deg C, a shoulder appears at the high island side of the distribution and transforms into a clearly separated second mode at higher temperatures. Hence, we find that bimodal distributions occur earlier (for lower GaN coverage) at higher substrate temperature. Another remarkable finding is that the height of the mode 1 islands remains constant as a function of temperature, whereas the height of mode 2 islands increases with temperature, again without signs of saturation in the examined temperature range.

\subsection{Discussion}

\begin{figure}[tb]
\includegraphics[width=8cm,clip]{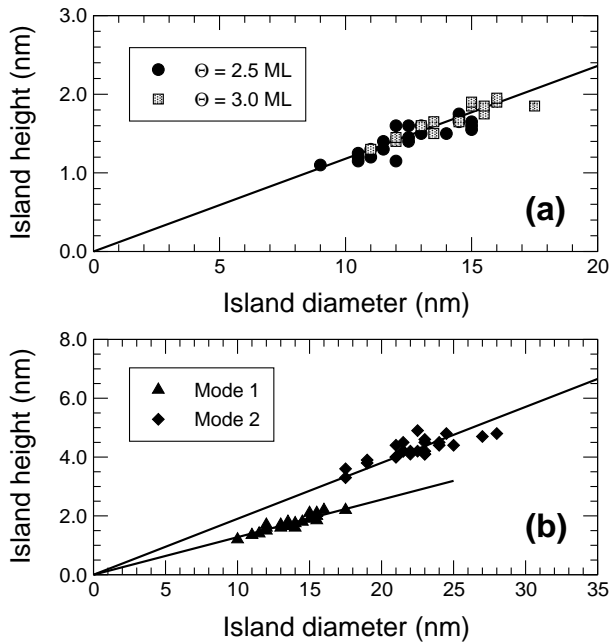}
\caption{\label{Fig:AspectRatio} \small  \textbf{(a)} Island height as a function of diameter for mode 1 islands obtained for two GaN coverages, as indicated. $T_S = 730$\,\deg C. The solid line corresponds to an aspect ratio of $\tau = 0.12$. \textbf{(b)} Island height as a function of diameter for mode 1 and mode 2 islands. $T_S = 730$\,\deg C, $\Theta = 4.0$\,ML. The solid lines correspond to aspect ratios of $\tau = 0.13$ for mode 1 and $\tau = 0.19$ for mode 2, respectively.}
\end{figure}

The above results demonstrate the occurrence of bimodal island size distributions at high GaN coverage and/or substrate temperature; at low GaN coverage and/or substrate temperature, the size distribution is monomodal. Remarkably, when bimodal distributions occur, the size of mode 1 islands (smaller islands) appears independent of growth parameters such as GaN coverage and substrate temperature. 

As discussed above, the variation of partial densities of both modes implies that first mode 1 islands are nucleated, their density saturates and the growth proceeds further by transformation of mode 1 islands into mode 2 islands. Thus the question arises, whether this transformation involves a shape change of the islands similar to the behavior observed in the Ge/Si system, where pyramidal islands are observed to transform into ``dome''-shaped islands during growth.\cite{Medeiros,Williams} Some information about the island shape can be gathered by RHEED. The RHEED images corresponding to pure mode 1 island morphologies and morphologies with a mixture of mode 1 and mode 2 islands (but with a majority of mode 2 islands) are both characterized by the same $\left\{10\bar{1}3\right\}$ facets, showing a six-fold rotation symmetry. It thus appears that both types of islands are (truncated) hexagonal pyramids with $\left\{10\bar{1}3\right\}$ sidewalls, in agreement with previous results.\cite{Daudin,Widmann} However, it is clear that the shape of such islands cannot be obtained from the RHEED pattern alone and further work is necessary to determine the precise shape of both types of islands.

A more detailed analysis can be done by extracting the aspect ratio $\tau$ of the islands from the AFM data. Figure \ref{Fig:AspectRatio}(a) depicts the aspect ratio of mode 1 islands for two different GaN coverages (at which monomodal height distributions are observed) at a substrate temperature of $730$\,\deg C. In spite of the absolute increase of the islands' size, their aspect ratio remains constant and we find $\tau = 0.12$. As shown in Fig.~\ref{Fig:AspectRatio}(b), the aspect ratio of mode 1 islands remains constant ($\tau = 0.13$) after the transition to a bimodal distribution but the mode 2 islands have a significantly larger aspect ratio of $\tau = 0.19$. Both aspect ratios correspond to truncated pyramids \cite{remark} but mode 1 islands are flatter.

\begin{figure}[tb]
\includegraphics[width=8cm,clip]{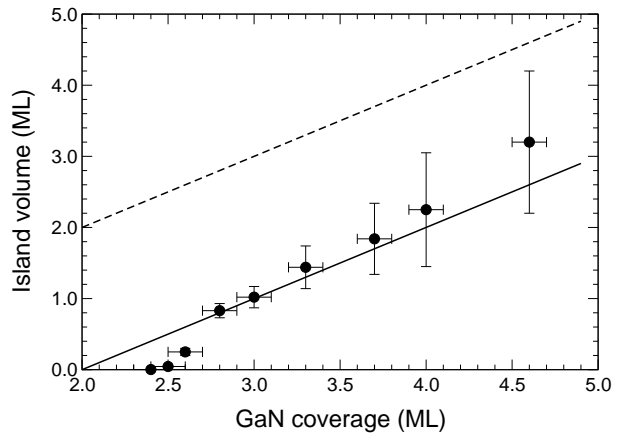}
\caption{\label{Fig:VolumefTheta} \small Total volume contained in 3D GaN islands as a function of coverage. The solid line indicates the expected behavior for a constant 2\,ML thick wetting layer, whereas the dashed line indicates the expected behavior for no wetting layer at all.}
\end{figure}

From the aspect ratio data, we can further calculate the amount of GaN contained in the 3D islands as a function of deposited GaN. The result is shown in Fig.~\ref{Fig:VolumefTheta}. We find that the data are consistent with a 2\,ML thick wetting layer, independent of the amount of deposited GaN. It thus appears that in this system, the wetting layer is stable and does not contribute to island growth. The deviation for GaN coverages just after the critical thickness of 2.25\,ML can be explained by the presence of 2D precursor islands, which still contain significant material but are not taken into account as 3D islands.

\section{Modeling}

In order to understand more precisely why the observed island distributions form, we analyze the nitride nanostructure growth in terms of a thermodynamic equilibrium model that has so far only been applied extensively to the Ge/Si system.\cite{Rudd} To date, little effort has been made to fit data on size distributions of nitride nanostructures to any model. Nakajima \emph{et al.}\cite{Nakajima} have undertaken a purely theoretical study and explicitly derived thickness-composition phase diagrams for the expected growth mode for GaInN/GaN and GaInN/AlN, using known materials parameters. However they did not take into account the shapes and sizes of experimentally observed nitride nanostructures, nor did they seek to predict size distributions. Here we take a rather different approach. We apply the thermodynamic model to the AFM data on nitride nanostructure distributions grown at one particular temperature in order to assess whether the bimodal distributions of nanostructures are consistent with thermodynamic equilibrium, as opposed to being configurations along an unstable ripening trajectory. The experimental distributions are described reasonably well by the model, as explained below. We then apply the model to understand the variations in the bimodal distributions with growth temperature at fixed coverage, and again find reasonable agreement over a range of growth temperatures. 

The thermodynamic equilibrium model makes explicit predictions for the form of the bimodal epitaxial nanostructure distributions.  The model is governed by a set of parameters that describes the size dependence of the internal energy of an epitaxial nanostructure.  Following Shchukin \emph{et al.},\cite{Shchukin} we describe the internal energy $\epsilon_i$ of the $i$th individual island of type $X$ and volume $\nu_i$ as follows: 

\begin{equation}
\epsilon_i = A_X\nu_i + B_X\nu_i^{2/3} + C_X\nu_i^{1/3} + D_X.
\label{Eq:Energy}
\end{equation}

\noindent Using this expression for the internal energy, the island size distribution $f(\nu_i)$ is given by

\begin{equation}
f(\nu_i) = \exp\left(- \frac{\epsilon_i - \mu \nu_i}{k_BT_S}   \right),
\label{Eq:Distribution}
\end{equation}

\noindent where $\mu$ is the chemical potential and $k_B$ is Boltzmann's constant.

Equation (\ref{Eq:Energy}) describes the chemical and elastic contributions to the island energy. The first term arises from bulk strain, the second from surface and interfacial energies and island-island interactions, and the third from surface stress and edge energies (suppressing the log dependence of the edge elastic relaxation energy). The equation also includes a size-independent term that is relevant only insofar as it differs from one nanostructure type to another. For the internal energy (\ref{Eq:Energy}), a minimum may exist if $B$ and $C$ have opposite signs. There has been some debate over which of the two constants is likely to be positive and which negative. In the initial development of a ``shape-map'' for the bimodal Ge/Si system by Williams \emph{et al.},\cite{Williams} $B$ was assumed to be negative and $C$ positive. In more recent work on Ge/Si, Rudd \emph{et al.}\cite{Rudd} have suggested that $B$ should be positive, since a Ge(001) surface does not spontaneously roughen, though they also acknowledge that, since 3D growth is preceded by the appearance of increasing numbers of defects in the wetting layer with their own associated energy, the behaviour of the Ge(001) surface of a bulk crystal may not be relevant and $B$ could be negative.  Here, we find that the option most consistent with the nanostructure size distributions observed by AFM is $B > 0$ and $C < 0$.  

Unlike in the most recent developments of the equilibrium model,\cite{Rudd} we have not, as yet, explicitly included the effects of the elastic interaction between pairs of islands. We also have not undertaken a self-consistent calculation of the chemical potential.  Instead, we have taken a similar approach to Williams \emph{et al.}:\cite{Williams}  Using the expressions for the distributions (\ref{Eq:Distribution}) and internal energies (\ref{Eq:Energy}), we have fitted the model to the available data at several coverages and temperatures, in order to determine approximately how the constants $A'$, $B$, $C$, and $D$ vary as a function of growth parameters.  For convenience, we have introduced $A' = A - \mu$. We then used our empirical functions to fit the model across multiple data sets, to determine how well it compares with the overall behaviour of the system. By comparison with the more sophisticated work of Rudd \emph{et al.}\cite{Rudd} on Ge/Si, we expect $B$ to be dependent on the coverage and $A'$ to depend on the chemical potential, a non-trivial function of coverage and temperature.        
  
\subsection{Evolution of size distribution with coverage}

In investigating the variation of the parameters $A'$, $B$, $C$, and $D$ as a function of coverage, we found that $C$ and $D$ could be treated as being independent of coverage, but that $B$ increased monotonically with increasing coverage. This coverage dependence is consistent with elastic island-island interactions, which are expected to scale with $\nu_i^{2/3}$, and which will increase with increasing coverage.  $B\left(\Theta\right)$ could be approximated to a straight line with non-zero slope for both the smaller ($S$) and the larger ($L$) island types.  Thus (for $X \in \left\{ S, L\right\}$):

\begin{equation}
B_X(\Theta ) \approx B_{0X} + b_X\Theta.
\end{equation}

Similarly, Williams \emph{et al.}\cite{Williams} found that, for the Ge/Si case, $C$ was independent of coverage and $B$ varied linearly with coverage.  However, they were also able to approximate the variation of $A'$ with coverage as a straight line with non-zero slope.  This does not appear to be valid in this case, and we have not found a simple functional relationship between $A'$ and coverage.  Hence in fitting the model across multiple data sets we have calculated a separate value of $A'$ for each data set, but have assumed a constant relationship between the value of $A'$ for the larger and smaller islands.  Essentially:

\begin{equation}
A'_{L}(\Theta) = g(\Theta) \quad \mathrm{and} \quad A'_{S}(\Theta) = g(\Theta) + A_0,
\end{equation}

\noindent where $A_0 = A_S - A_L$, representing the intrinsic differences in bulk elastic energy between the two island types.  The function $g(\Theta)$, which has not been found explicitly, represents the coverage dependence of the chemical potential.    

\begin{figure}[tb]
\includegraphics[width=8cm,clip]{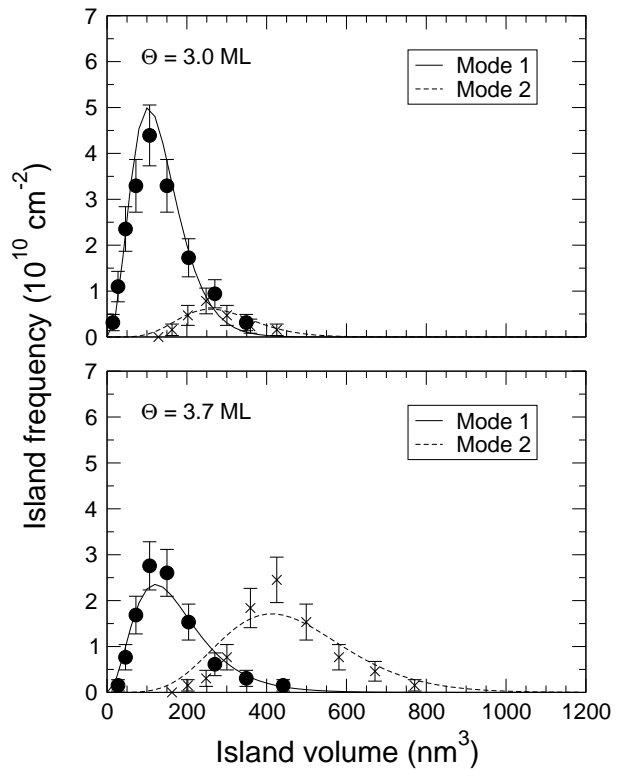}
\caption{\label{Fig:R1} \small Comparison of experimentally observed island size distributions at 730\,\deg C with fitted island size distributions for both the smaller (circles) and larger (crosses) islands, for coverages of 3.0\,ML and 3.7\,ML, as indicated. The fitting has been done, as described in the text, with a single set of parameters to describe a range of coverages. Values of the $C$- and $D$-parameters do not vary with coverage, values of the $B$-parameters vary linearly with coverage and values of the $A$-parameter are individually fitted such that $A_S - A_L = A_0$ for all coverages.}
\end{figure}

Hence, in fitting across four coverages at which large and small islands coexist, we used the parameters $B_{0S}$, $B_{0L}$, $b_S$, $b_L$, $C_S$, $C_L$, $D_S$, $D_L$, $A_0$, and four separately fitted values of $g(\Theta)$.  This clearly gives us a large number of parameters---however, the model must fit a large number of observable features of the data.  We are able to provide a reasonable fit for each shape for the relative height, position, width and skewness of the volume distributions (essentially 15 experimental observables for each island shape).  Example fits are shown in Fig. \ref{Fig:R1}.  The fitting has been done using a $\chi^2$ approach and the goodness-of-fit may be judged by calculating an overall $Q$-value across the data sets, $Q_\mathrm{total}$.  For the data sets under consideration, $Q_\mathrm{total} = 0.22$.  Since the available data sets are much smaller for this system than for the Ge/Si system (in which counting large numbers of islands is facilitated by the larger sizes of the islands, and the clear shape difference between island types) this $Q$-value is satisfactory.  For comparison we have also attempted to fit the data with the original model of Williams \emph{et al.},\cite{Williams} and with some Gaussian functions.  These fits were markedly worse and it was not possible to find parameters that fit the entire set of data with a reasonable $Q$-value.  From these findings we may infer that the data are consistent with the equilibrium model; however, the model can in no way be said to have been proven. The model contains a large number of parameters for which little \emph{a priori} information is available in the nitride systems, and the measured island distributions would benefit from more data in order to reduce the statistical uncertainty.  Hence, the discussion that follows is somewhat tentative.

\begin{figure}[tb]
\includegraphics[width=8cm,clip]{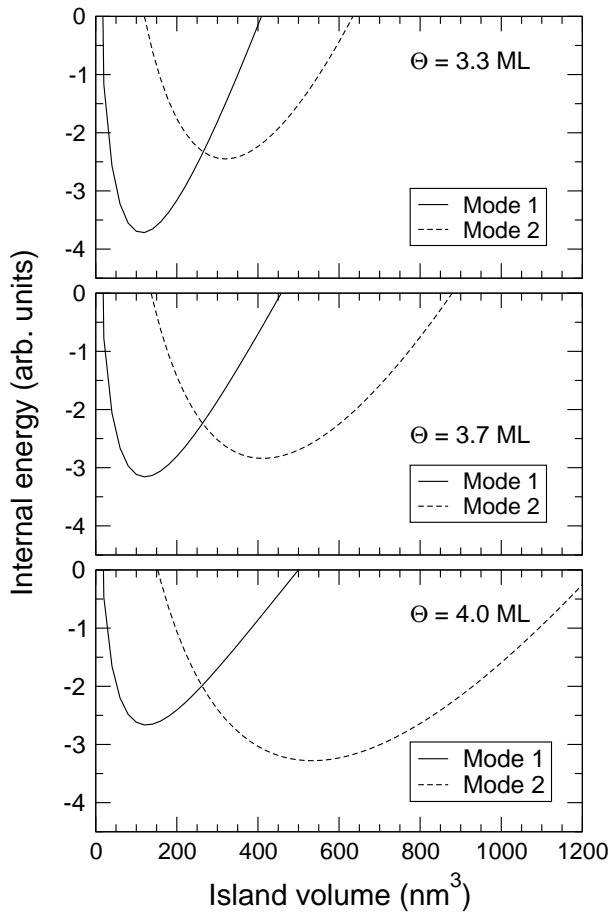}
\caption{\label{Fig:R2} \small Plots of the variation in internal energy of an island with island volume for the smaller (solid line) and larger (dashed line) islands at 730\,\deg C for three coverages, as indicated.}
\end{figure}

Using the calculated parameters we can plot the internal free energy versus volume curves for both the smaller and the larger islands.  The results are shown in Fig.~\ref{Fig:R2} for three different coverages. There is a significant evolution of the curves as the coverage increases. Similar results were observed by Williams \emph{et al.}\cite{Williams} for the evolution of internal energy with coverage in the Ge/Si system, and were attributed to the interactions of the nanocrystals.  The value of $B(\Theta)$ affects the width of the curve and the horizontal position of the energy minimum. In the Ge/Si case, Williams \emph{et al.} found that for the larger islands (domes) the value of $B(\Theta)$ increases (becomes less negative) with increasing coverage, whilst the value of $B(\Theta)$ for the smaller islands (pyramids) decreases, which suggested that the repulsion between domes was stronger than the repulsion between pyramids. In contrast, the work of Rudd \emph{et al.}\cite{Rudd} (also on Ge/Si) suggests that $B(\Theta)$ increases with increasing coverage at a similar rate for both pyramids and domes.  Similarly, in the current case, we find that the $B$-parameter increases with increasing coverage at almost the same rate for both island types, indicating no significant difference between the elastic island-island interactions for each island type.  

Changes in $A'(\Theta)$ tend to shift the internal-free energy versus volume curves vertically. At low coverages, the minimum internal energy is lower for the smaller islands, but at high coverages, it is lower for the larger islands.  Hence the formation of large islands rather than small islands is preferred at high coverages.  However, the elastic interactions between islands may mean that not all small islands grow to become large islands.  Some may be destabilised by neighbouring islands and dissolve.  One surprising feature of the fits is that $A'$ is found to be positive for the larger islands, for all coverages. Since the bulk elastic energy is expected to decrease when an island is formed, this suggests that the chemical potential is negative and relatively large (since $A' = A - \mu$).  This finding is not yet well understood.

\subsection{Evolution of size distribution with temperature}

In order to understand the variation of the size distribution with temperature, we must consider the temperature dependence of the free energy of each island type.  In principle, $A'$, $B$, and $C$ may all be temperature dependent. The size-independent $D$-parameter is constant and is found to be the same for both small and large islands.  Williams \emph{et al.}\cite{Williams} performed some limited studies on the temperature dependence of the fitting parameters in Ge/Si and found that only $A'$ exhibited significant temperature dependence. Similarly, in this case, $C$ is found to be essentially independent of temperature.  However, for both island types, we have found that $B$ exhibits a significant temperature dependence, such that for $X \in \left\{S, L\right\}$:

\begin{equation}
B_X(T_S) = B_{1X} + \beta_XT_S.
\end{equation}

The $B$-parameter increases at a similar rate for the larger and the smaller islands, implying that the surface energy contribution from each island type varies in a similar way with temperature. This is congruent with the suggestion that the $\left\{10\bar{1}3\right\}$ facets seen in RHEED dominate both island types.  These results are not, however, incompatible with the existence of other facets on the larger island types.      

The $A'$-parameter, which was found to exhibit a complex dependence on coverage, also exhibits a complex dependence on temperature.  Hence in fitting the model across a range of temperatures, we have calculated values of $A'$ individually for each data set, but have assumed a constant relationship between the value of $A'$ for the large islands and the value of $A'$ for the small islands, such that

\begin{equation}
A'_{L}(T_S) = h(T_S) \quad \mathrm{and} \quad A'_{S}(T_S) = h(T_S) + A_1.
\end{equation}

\begin{figure}[tb]
\includegraphics[width=8cm,clip]{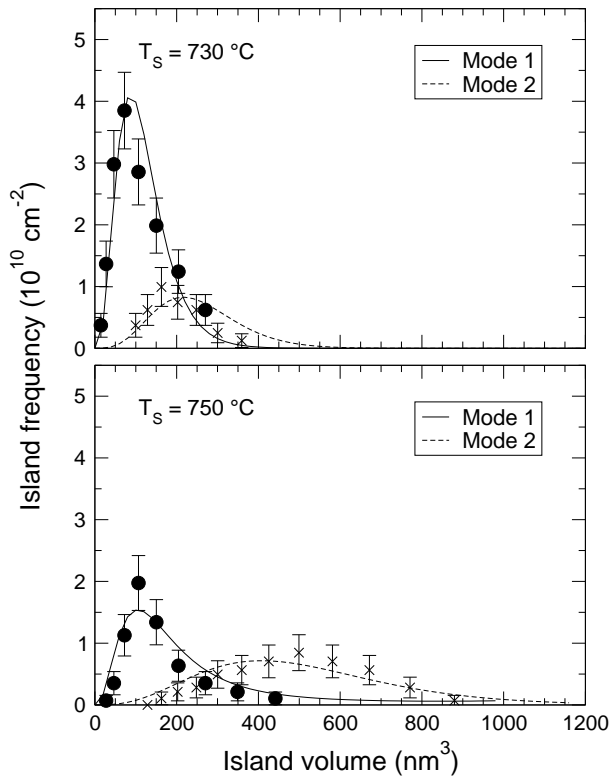}
\caption{\label{Fig:R3} \small Comparison of experimentally observed island size distributions with fitted island size distributions for both the smaller (circles) and larger (crosses) islands, for a coverage of 3.0 ML and temperatures 730\,\deg C and 750\,\deg C, as indicated. The fitting has been done, as described in the text, with a single set of parameters to describe a range of temperatures. Values of the $C$- and $D$-parameters do not vary with temperature, values of the $B$-parameter vary linearly with temperature and values of the $A$-parameter are individually fitted such that $A_S - A_L = A_1$ for all coverages.}
\end{figure}

For the larger islands, $A'$ is again found to be positive at lower temperatures but decreases smoothly with temperature and changes sign at $\approx 755$\,\deg C.  The fits are shown in Fig.~\ref{Fig:R3}; $Q_\mathrm{total} = 0.73$.  The function $h(T_S)$, which has not been found explicitly, includes the temperature dependence of the chemical potential, and any variation in the elastic relaxation with temperature.

This model is only valid if a quasi-equilibrium may be assumed.  This assumption is particularly problematic with regards to variation in temperature, since in particular at lower temperatures, the system may not reach equilibrium, whilst at higher temperatures significant interdiffusion between the islands and the substrate may occur. This alloying will effectively reduce the mismatch between the islands and the substrate - strongly affecting the island size distributions, since the mean equilibrium nanostructure volume varies inversely as the sixth power of the mismatch.\cite{Dorsch}  Rudd \emph{et al.}'s more sophisticated work,\cite{Rudd} in creating the nanostructure diagram for the Ge/Si system, could be augmented to account for uniformly distributed alloying, and for kinetic effects. In the past, whilst the temperature dependence of the equilibrium island size distribution has been considered theoretically,\cite{Rudd,Williams} very little high quality data has been available for comparison.  The experimental studies by Williams \emph{et al.}\cite{Williams} only involved temperatures of 550\,\deg C, 600\,\deg C, and 650\,\deg C, and there was considerable uncertainty in the temperature measurement. Additionally, the sample grown at 550\,\deg C is believed to have been significantly influenced by kinetic effects, and the sample grown at 650\,\deg C experienced significant intermixing.  Hence the data considered here, relating to a smaller range of more accurately measured temperatures, are very valuable as a first quantitative test of the temperature dependence of the equilibrium model. In addition, the GaN/AlN system presents an important advantage over the Ge/Si system since it has been show that alloying is very limited even at high growth temperatures.\cite{Arlery}

In considering the results of the fitting, we notice that fitting across the temperature data sets gives a higher $Q_\mathrm{total}$ than fitting across the coverage data, with an essentially similar fitting methodology.  One might thus speculate that, since the samples used for the coverage data set were grown at the relatively low temperature of 730\,\deg C, full equilibrium has not been achieved and kinetic effects are having some influence.  A worthwhile approach, in investigating this further, will be to examine a series of samples that have been annealed at growth temperature to allow further evolution towards equilibrium.

\section{Conclusion}

Using \emph{ex situ} AFM, we have studied the evolution of thin GaN layers grown by plasma-assisted MBE on AlN(0001) layers at substrate temperatures between 700 and 750\,\deg C. Initially, 2\,ML of GaN grow in a 2D mode, followed by the occurrence of 2D islands. These islands act as precursors for 3D islands, which appear after an SK transition around 2.3\,ML. During further growth, in particular at higher temperatures, a bimodal island size distribution is observed. Remarkably, the size of mode 1 islands is found to be independent of coverage and temperature, whereas the size of mode 2 islands increases with coverage and temperature. The analysis of the partial island densities reveals that, whilst the total island density remains constant, mode 1 islands transform during growth into mode 2 islands. The aspect ratio of the islands is measured for both types of islands and it is found that they are characterized by distinctively different aspect ratios, whereas no additional facets are observed in the RHEED pattern due to mode 2 islands.

These findings are examined in the framework of an equilibrium model for SK growth.  We find satisfactory agreement with the experimental data, suggesting that, as more data become available, extending the equilibrium approach may be helpful in understanding and tailoring nitride nanostructure distributions.  We have examined the variation of each of the fitting parameters with growth conditions, and have considered how this may relate to the physics of this system.  The calculated parameters appear to be compatible with the available data---for example the $B$-parameter, relating to the surface energy of the islands shows a similar variation with growth parameters for each island type, which is unsurprising if both island types are dominated by the same facet.  However, the fits are in no way perfect, and this may be due to the influence of kinetic effects on the island distributions.  Further experiments on the annealing of dot arrays at growth temperature might clarify to what extent the growth proceeds near equilibrium or kinetic effects are predominant.

\begin{acknowledgments}

The authors would like to thank O. Briot (Universit\'e Montpellier II, France) for providing the GaN templates and H. Mariette (Universit\'e Joseph Fourier, Grenoble, France) for many stimulating discussions. R.E.R.'s contribution to this work was performed under the auspices of the U.S. Department of Energy by the University of California, Lawrence Livermore National Laboratory, under Contract No.~W-7405-Eng-48.

\end{acknowledgments}


\begin{thebibliography}{99}
\bibitem{remark} The aspect ratio for a full pyramid with $\left\{10\bar{1}3\right\}$ facets is $\tau = 0.29$.
\end{thebibliography}
\end{document}